# A new route to spin-orbit torque engineering via oxygen manipulation

Xuepeng Qiu[1], Kulothungasagaran Narayanapillai[1], Yang Wu[1], Praveen Deorani[1], Xinmao Yin[2], Andrivo Rusydi[2], Kyung-Jin Lee[3,4], Hyun-Woo Lee[5], and Hyunsoo Yang[1,*]

[1]*Department of Electrical and Computer Engineering, and NUSNNI, National University of Singapore, 117576, Singapore*

[2]*Singapore Synchrotron Light Source, Department of Physics, and NUSNNI, National University of Singapore, 117603, Singapore*

[3]*Department of Materials Science and Engineering, Korea University, Seoul 136-701, Korea*

[4]*KU-KIST Graduate School of Converging Science and Technology, Korea University, Seoul 136-713, Korea*

[5]*PCTP and Department of Physics, Pohang University of Science and Technology, Kyungbuk 790-784, Korea*

*e-mail: eleyang@nus.edu.sg

**Spin transfer torques allow for electrical manipulation of magnetization at room temperature, which is utilized to build future electronic devices such as spin transfer torque memories. Recent experiments have discovered that the combination of the spin transfer torque with the spin Hall effect enables more efficient manipulation. A versatile control mechanism of such spin-orbit torques is beneficial to envision device applications with competitive advantages over the existing schemes. Here we report that the oxidation**

**manipulation of spin-orbit torque devices triggers a new mechanism, and the resulting torques are estimated to be about two times stronger than that of the spin Hall effect. Our result introduces an entirely new way to engineer the spin-orbit torques for device operation via oxygen manipulation. Combined with electrical gating for the control of the oxygen content, our finding may also pave the way for towards reconfigurable logic devices.**

Controlling the magnetization direction via interaction between spins and charges is crucial for spintronic memory and logic devices[1-4]. Conventional magnetization switching using current-induced spin transfer torques (STTs) requires a spin polarizer in a spin valve structure[5,6]. Recently, the combination of STTs with the spin Hall effect has led to a new type of torque, namely spin-orbit torques (SOTs). In magnetic bilayers, where an ultrathin ferromagnetic (FM) layer is in contact with a heavy metal (HM) layer, SOTs allow the manipulation of the magnetization by an in-plane current[7-12]. SOTs enable low power magnetization switching[7,8], fast domain wall motion[9,13,14], and tunable nano-oscillators[11,12]. The high stability, simplicity, and scalability of the scheme make it attractive for next-generation spintronic devices[8,15]. Therefore, further control of the magnitude and even sign of the SOTs is beneficial to fully utilize the envisioned SOT devices.

So far, the scope in the experimental studies of SOT has been limited either to the HM material replacement[8,14,16] using Pt, Ta, and W, or to the thickness variation[17-19] of the HM or FM layer. The replacement of Pt by Ta reverses the sign of the SOTs[8,14], which is attributed to the opposite signs of the spin Hall angle in these two materials. The sign change of the SOTs is suggested as a reason for the opposite directions of the current-driven domain wall motion in Pt-based and Ta-based nanowires[14]. As the SOTs can possibly be altered via interface engineering, especially at the HM/FM interface, we examine the oxygen bonding effect on SOTs in Pt/CoFeB/MgO/$SiO_2$. Since the Pt layer, which is the source of the bulk spin Hall effect, is hardly affected by oxygen in our experiment, any change of the measured SOTs due to oxygen bonding should be attributed to a different mechanism arising from either the CoFeB layer or the Pt/CoFeB interface. We find that as the oxygen bonding level increases, not only a magnitude change but also a full sign reversal of SOTs occur. While the magnitude change can be explained within the bulk spin Hall model, the full sign reversal goes beyond the bulk spin Hall effect and evidences a new SOT mechanism. We estimate that the new mechanism in our sample can be two times stronger than the bulk spin Hall mechanism, resulting in greater tunability of SOTs compared with the bulk spin Hall effect proposed previously.

Since MgO is hygroscopic[20], the oxygen bonding level can be controlled by the thickness of the $SiO_2$ capping layer. The sputter-deposited film structure of Pt/CoFeB/MgO/$SiO_2$ is shown in Fig. 1a, in which the thickness $t$ of the $SiO_2$ layer is varied from 0 to 5 nm. A scanning electron micrograph (SEM) image of the patterned Hall bar is shown in Fig. 1b. We find that $t$ variation alters SOTs considerably. Figure 1c shows the anomalous Hall resistance ($R_H$) of the device as a function of the in-plane current ($I$) applied to the device. In addition to the current, a small constant magnetic field of 400 Oe is applied along the current direction to break the symmetry[7,8] of the device and allow for selective magnetization switching by the in-plane current. Since $R_H$ probes the average $z$-component of the CoFeB magnetization, the hysteretic switching of $R_H$ confirms that the current-induced SOTs indeed switch the magnetization. Interestingly the switching sequence, marked by arrows, is clockwise for $t > 1.5$ nm but counterclockwise for $t \leq$ 1.5 nm, for which we expect significant oxygen bonding. Only the switching sequence for large $t$ (low oxygen level) is consistent with the previously reported switching sequence for Pt/Co/AlOx[7,21] and Pt/CoFe/MgO[14]. Thus, the switching sequence for small $t$ is abnormal. The current-induced Oersted field cannot explain the sequence reversal. The sequence reversal is not due to the sign reversal of the relation between $R_H$ and the $z$-component of the magnetization either, since the purely magnetic-field driven magnetization switching (see Supplementary Information) does not exhibit the switching sequence reversal with $t$. The inset of Fig. 1e shows $I_S$, which is defined as the current at which $R_H$ changes from positive to negative. Note the sign change of $I_S$ around $t$ = 1.5 nm. Figure 1e shows the switching current $I_S$ divided by the anisotropy field $H_{an}$ as a function of $t$. Note that the ratio maintains roughly the same magnitude before and after the sign reversal point. Since the ratio ($I_S/H_{an}$) provides a rough magnitude estimation for SOTs[15,21], it implies that SOTs remain strong even after the sign reversal, which is in contrast to weak sign reversal in a previous study[18]. Thus the contribution of the new SOT mechanism, which is triggered by the oxygen bonding, should be about two times larger than the bulk spin Hall contribution and of opposite sign.

For an independent test of SOT sign reversal, we perform lock-in technique measurements of SOTs[18,22]. We apply a small amplitude sinusoidal ac current with a frequency of 613.7 Hz to exert periodic SOTs on the magnetization, so that the induced magnetization oscillation around the equilibrium direction generates the second harmonic signal $V_{2\omega}$. Depending on the measurement geometry, the damping-like SOT or field-like SOT, which are two mutually orthogonal vector components of SOTs, are measured[18,23]. As current induced magnetization switching is driven mainly by the damping-like SOT[7,8,21], we present here the result for the damping-like SOT, which can be probed by applying an external dc magnetic field $H$ along the current direction (with 1 degree tilting angle from the film plane) to tilt the equilibrium magnetization direction accordingly. The result for the field-like SOT is presented in the Supplementary Information, and also shows a sign reversal like the damping-like torque, as described below. Asymmetric $V_{2\omega}$ loops have been observed as shown in Fig. 1d. For $t$ = 0 and 1.2 nm, there is a dip at a positive field and a peak at a negative field, whereas an opposite behavior is observed in the cases of $t$ = 2.1 and 3 nm. Opposite polarities in $V_{2\omega}$ indicate that the damping-like SOT is pointing in opposite directions for the thinner and the thicker cases, confirming the conclusion drawn from the switching sequence in Fig. 1c. When $t$ = 1.5 nm, the $V_{2\omega}$ signal for positive field contains both a peak and a dip, which may indicate the coexistence of regions with opposite damping-like SOT directions. The peak or dip values ($V_{2\omega-\max}$) at positive magnetic fields are summarized in Fig. 1f for various $t$. The peak and dip heights are comparable, indicating the full sign reversal of damping-like SOT in agreement with Fig. 1e.

In order to verify that oxygen indeed penetrates the magnetic bilayer when $t$ is small, we have carried out four independent experimental studies, such as secondary ion mass spectroscopy (SIMS), X-ray magnetic circular dichroism (XMCD), X-ray absorption spectra (XAS), and X-ray photoelectron spectroscopy (XPS). Figure 2a provides the oxygen depth profile obtained by SIMS for the two devices with $t$ = 0 and 2 nm. The depth profile for $t$ = 0 nm shows a significantly enhanced oxygen level in the CoFeB layer compared to that for $t$ = 2 nm, confirming excess oxygen for small $t$. On the other hand, the two depth profiles are almost identical in the Pt layer. The Pt 4$f$ XPS spectra in Fig. 2b also prove that the oxygen penetration through the thin $SiO_2$ capping layer has not affected the Pt layer, which is natural since Pt has an excellent resistance to oxidation. XMCD probes, on the other hand, magnetic properties of specific elements[24]. Both the Co (Fig. 2c) and Fe (Fig. 2d) $L_{3,2}$ edge signals are much weaker for $t$ = 0 nm than for $t$ = 3 nm. Since the oxidation of the magnetic elements reduces their magnetic moments, the XMCD result is consistent with the oxidation of the FM layer for small *t*. The XAS data in Fig. 2e,f indicate the excess oxygen bonding to Fe rather than Co in the case of $t = 0$.[25]

Although all data so far are consistent with the oxidation of devices, we cannot yet rule out effects of other *t*-dependent properties such as lattice tension relaxation. To rule out such possibilities, we change the oxygen level *without* changing *t*. We achieve this by varying temperature or vacuum level. For this purpose, we focus on a *single* sample with the critical thickness $t = 1.5$ nm. At normal vacuum level (8 Torr), the current-induced switching sequence shows a normal clockwise behavior at low temperature (10 K) but reverses to abnormal anti-clockwise behavior at high temperature (200 K) as shown in Fig. 3a. Since the anomalous Hall and the planar Hall effects do not change sign with temperature (see Supplementary Information), the switching sequence reversal implies the temperature-induced sign reversal of the damping-like SOT. Since a lower (higher) oxygen level is expected at lower (higher) temperature, the abnormal switching sequence at higher temperature is consistent with the oxygen bonding effect. The complex shape of the switching sequence at intermediate temperature is not clearly understood but it may be connected with the origin of the unusual $V_{2\omega}$ line shape (Fig. 1d) near the SOT reversal point.

Another test is the vacuum level dependence shown in Fig. 3. When the same experiment is performed at higher vacuum (~$10^{-5}$ Torr) in Fig. 3b, the switching sequence reversal from normal clockwise to abnormal anti-clockwise direction is systematically deferred to higher temperatures. Since a higher vacuum level is expected to suppress the oxygen level within the sample, this experimental data also support the oxygen bonding effect on SOT.

We now discuss mechanisms of SOTs. The bulk spin Hall effect[8] in the HM layer is an important mechanism for SOTs. According to the spin Hall theory of SOTs[14,21], the sign of the damping-like SOT is determined by the bulk spin Hall angle of the HM layer. This theory is thus inadequate to explain the oxygen-induced sign reversal of the damping-like SOT, since the HM layer is not affected by oxidation. Thus, our data necessitate an additional or new mechanism of SOTs arising from either the FM layer or the HM/FM interface. These two possibilities are hard to distinguish in our experiment, since the CoFeB FM layer is only 0.8 nm thick. One possible mechanism is the interfacial spin-orbit coupling[26-30] contribution to the damping-like SOT. The Rashba-like interfacial spin-orbit coupling[28,31] arises from the inversion asymmetry at the HM/FM interface, and it is plausible that oxygen may strengthen the asymmetry and thereby enhances the interfacial coupling strength as exemplified by the enhanced Rashba coupling at the magnetic Gd surface due to oxidation[32]. It was recently suggested[33] that the Berry phase effect makes the interfacial spin-orbit coupling contribution to the damping-like torque much larger than previous theories[26-30] predicted. Combined with the possible enhancement of the structural inversion asymmetry at the interface due to oxidation, this provides a possible explanation of our experimental results. Another possible mechanism is the interplay of the bulk spin Hall effect and the electronic structure of the FM layer through oxygen. A recent first principles calculation[34] compares the damping-like SOT in Pt/Co bilayer and Pt/Co/O trilayer, and finds that the spin Hall contribution to the damping-like SOT may vary considerably, when a FM layer is in contact with an oxygen layer as in Pt/Co/O. When oxygen atoms penetrate into a FM layer as in our experiment, the effect of oxygen may be stronger, providing another explanation for our experiment.

In order to further elucidate the effect of oxygen, we have measured the channel resistance and the Hall resistance as a function of temperature. In our experimental setup, the channel resistance represents mostly the property of the Pt layer, since the other metal layer (CoFeB) is only 0.8 nm thick and the resistivity of CoFeB is higher than that of Pt. Figure 4 shows that the temperature dependence of the channel resistance is similar regardless of *t*, suggesting no noticeable change in Pt. On the other hand, the anomalous Hall resistance probes mainly the CoFeB layer, since non-magnetic layers cannot contribute to the anomalous Hall effect. Even when Pt acquires induced magnetic moment[13] due to the proximity effect with CoFeB, its contribution to the anomalous Hall effect is weaker than that of CoFeB, since the proximity-induced magnetism is weaker than the magnetism of CoFeB. Note that the temperature dependence of the anomalous Hall effect shows two distinct behaviors depending on *t*. When $t \leq 1.5$ nm, the anomalous Hall resistance decreases with increasing temperature. In contrast, the anomalous Hall resistance shows a nonmonotonic behavior with temperature for $t >$ 1.5 nm. Although it is difficult to deduce a main mechanism of this *t*-dependent dramatic change because of the relatively narrow range of Hall resistance variation, these data indicate that FM layers for thinner and thicker samples are qualitatively different, consistent with the oxidation of the FM in samples with thinner $SiO_2$ capping.

Our result reveals that oxygen bonding to the FM activates another SOT mechanism, which arises from either the FM layer or the HM/FM interface. The oxygen bonding related mechanism may be combined with the spin Hall mechanism to design and control the switching direction due to SOTs, as well as enhance the efficiency for device applications. It may also be relevant for recent experimental controversies. The recent data from various groups show controversial results. Our result indicates that even for exactly the same layer structure, very different SOTs can be obtained depending on the details of the device preparation procedures, which may affect oxygen content in the sample. This provides a new way to investigate SOTs and paves the way for ionic engineering of SOTs, since the property of the materials can be modified during the sample growth or treatment in different environments. Two recent experiments[35,36] demonstrated that the degree of oxygen bonding can be electrically controlled via gating. A combination with this gating may augment the functionalities of SOT devices to allow, for instance, reconfigurable logic operation.

**Methods**

The stacked films of thermally oxidized Si substrate/MgO (2)/Pt (2)/$Co_{60}Fe_{20}B_{20}$ (0.8)/MgO (2)/$SiO_2$ (*t*) (numbers are nominal thicknesses in nanometers) with various $SiO_2$ capping thicknesses ($t = 0 \sim 5$) were deposited by magnetron sputtering with a base pressure $< 2 \times 10^{-9}$ Torr at room temperature as shown in Fig. 1a. The bottom MgO layer is used to promote perpendicular anisotropy. After deposition, the films were post-annealed at 300 °C for 1 hour in a vacuum to obtain perpendicular anisotropy. The multilayers were coated with a ma-N 2401 negative e-beam resist and patterned into 600 nm width Hall bars by electron beam lithography and Ar ion milling as shown in Fig. 1b. PG remover and acetone were used to lift-off the e-beam resist. Contact pads were defined by photolithography followed by the deposition of Ta (5 nm)/Cu (150 nm)/Ru (5 nm) which are connected to the Hall bars. Before the deposition of contact pads, Ar ion milling was employed to remove the $SiO_2$ and partial MgO layers to make low-resistance electric contacts. All the devices were processed at the same time to ensure the same fabrication processes, which is important for this study. Devices were wire-bonded to the sample holder and installed in a physical property measurement system (Quantum Design).

We performed the measurements of current induced switching and the second harmonic anomalous Hall voltage loops at 200 K for the data set in Fig. 1, at which all the devices with different $SiO_2$ capping thicknesses retain desirable perpendicular anisotropy (Supplementary

Information). A combination of Keithley 6221 and 2182A is used to apply a pulsed DC current of a duration of 50 μs to the nanowires and measure the Hall voltage, simultaneously. The pulsed DC current with an interval of 0.1 s is used to eliminate the accumulated Joule heating effect.

## Acknowledgments

This work was supported by the Singapore NRF CRP Award No. NRF-CRP 4-2008-06. K.L. acknowledges financial support by the NRF (NRF-2013R1A2A2A01013188).

## Author contributions

H.Y. and X.Q. planned the study. X.Q. and K.N. fabricated devices. X.Q. measured transport properties. Y.W. helped characterization. A.R. and X.Y. carried out x-ray measurements. All authors discussed the results. X.Q., K.L., H.L., and H.Y. wrote the manuscript. H.Y. supervised the project.

## Figure captions

**Fig. 1. Device structure and effect of capping layer.** **a**, Film structure of the multilayers. **b**, Scanning electron micrograph of the device. **c**, $R_H$ as a function of pulsed currents for different *t* at 200 K. *t* is indicated in each graph. A fixed 400 Oe magnetic field is applied along the positive current direction. The width of pulsed currents is 50 μs, and $R_H$ is measured after a 16 μs delay. **d**, Second harmonic anomalous Hall voltage $V_{2\omega}$ as a function of magnetic field (*H*). *H* is applied along the current direction with a tile of 1° away from the film plane. **e**, Switching current $I_S$ (defined as the switching current from a high to low $R_H$) divided by anisotropy field $H_{\text{an}}$ as a function of *t*. The inset shows $I_S$. **f**, The peak (dip) value of second harmonic loops $V_{2\omega-\max}$ at positive magnetic fields with different *t*.

**Fig. 2. SIMS, XPS, XMCD, and XAS characterization.** **a**, SIMS depth profiles of oxygen for the films without (*t* = 0 nm) and with (*t* = 2 nm) $SiO_2$ capping. Expected layers with the etching depth are written on the bottom of the graph. **b**, XPS spectra of Pt for the films without (*t* = 0 nm) and with (*t* = 2 nm) $SiO_2$ capping. **c**,**d**, XMCD of Co (**c**) and Fe (**d**) without (*t* = 0 nm) and with (*t* = 3 nm) $SiO_2$ capping. **e**,**f**, XAS of Co (**e**) and Fe (**f**) without (*t* = 0 nm) and with (*t* = 3 nm) $SiO_2$ capping.

**Fig. 3. Temperature and vacuum effect on current-induced switching.** Current-induced switching at different temperatures for the device with the critical thickness, *t* = 1.5 nm under normal (**a**: 8 Torr) and high vacuum (**b**: ~$10^{-5}$ Torr) conditions.

**Fig. 4. Temperature dependence of $R_{\text{channel}}$ and $R_{\text{H}}$ for the devices with different *t*.**

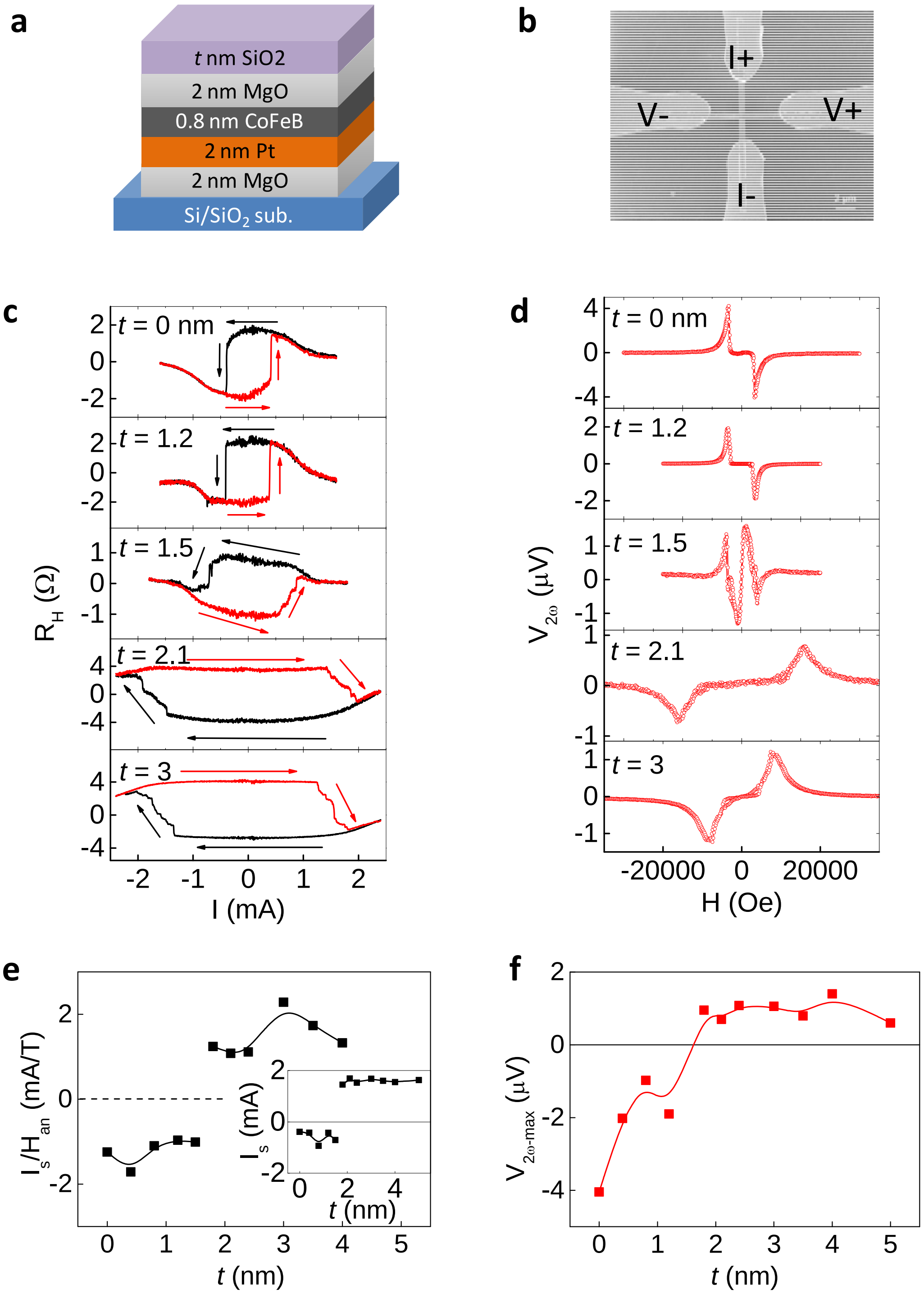

Fig. 1

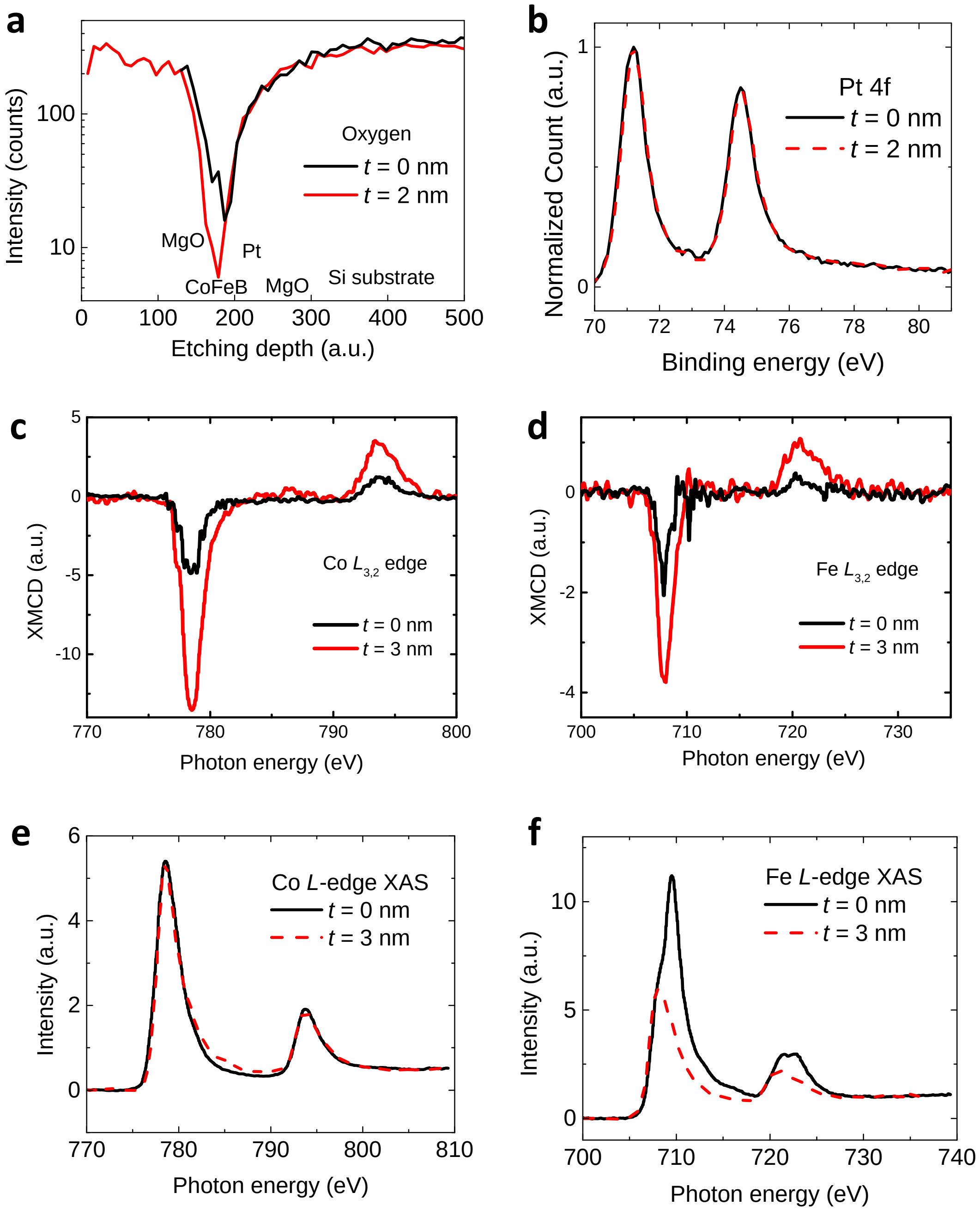
a
Intensity (counts)
Etching depth (a.u.)
Oxygen
t = 0 nm
t = 2 nm
MgO
CoFeB
Pt
MgO
Si substrate
b
Normalized Count (a.u.)
Binding energy (eV)
Pt 4f
t = 0 nm
t = 2 nm
c
XMCD (a.u.)
Photon energy (eV)
Co L3,2 edge
t = 0 nm
t = 3 nm
d
XMCD (a.u.)
Photon energy (eV)
Fe L3,2 edge
t = 0 nm
t = 3 nm
e
Intensity (a.u.)
Photon energy (eV)
Co L-edge XAS
t = 0 nm
t = 3 nm
f
Intensity (a.u.)
Photon energy (eV)
Fe L-edge XAS
t = 0 nm
t = 3 nm

Fig. 2

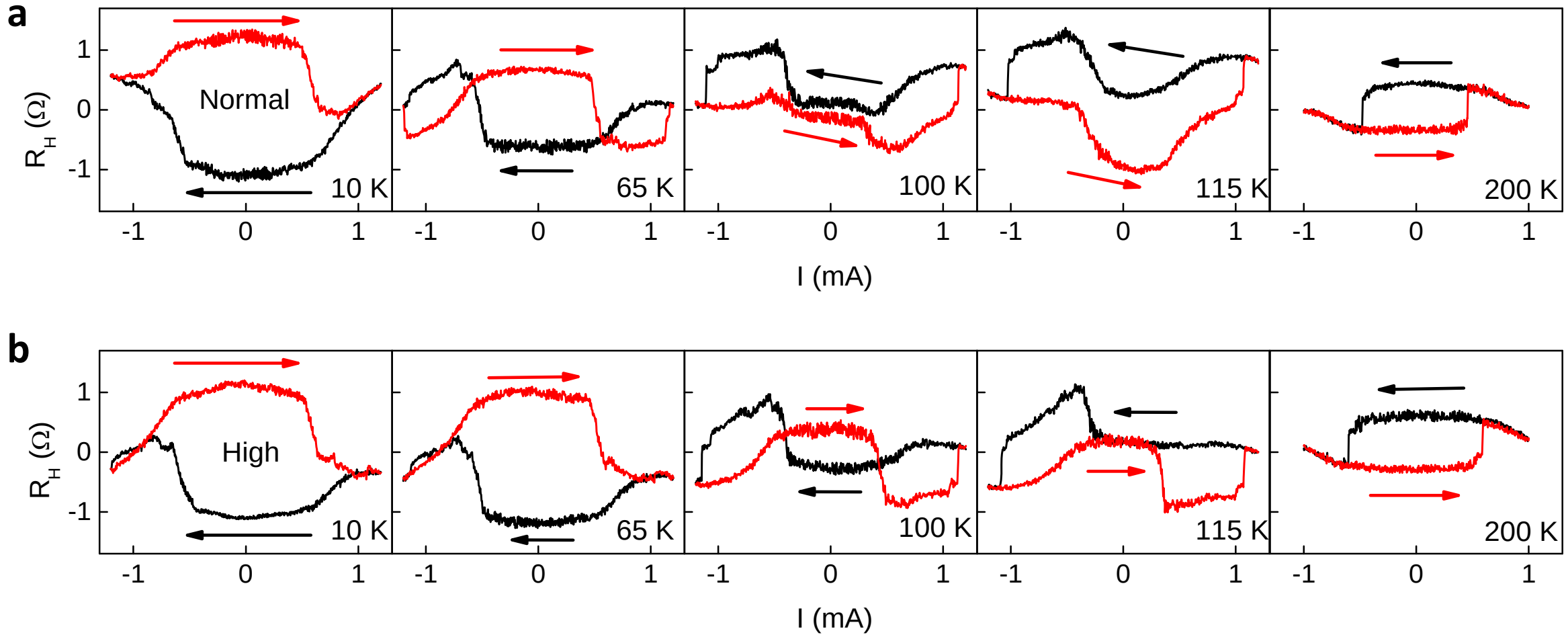

Fig. 3

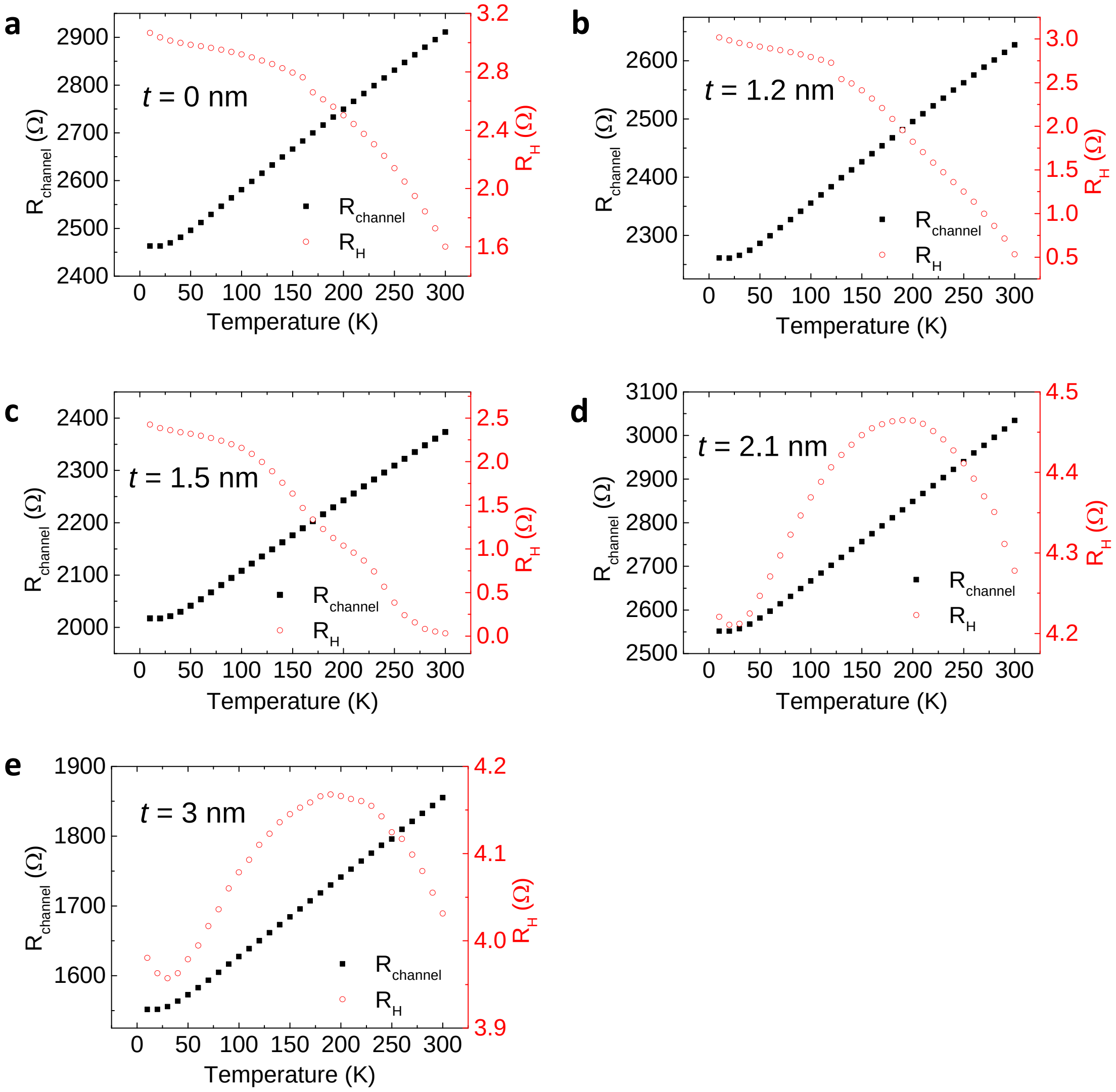
a
t = 0 nm
b
t = 1.2 nm
c
t = 1.5 nm
d
t = 2.1 nm
e
t = 3 nm
R_channel (Ω)
R_H (Ω)
Temperature (K)
R_channel
R_H

Fig. 4